\documentclass[prl,twocolumn,showpacs,superscriptaddress]{revtex4}

\usepackage{graphicx}
\usepackage{dcolumn}
\usepackage{bm}

\begin{document}

\preprint{Phys. Rev. Lett.}

\title{Constraints on Cosmological Parameters from the Analysis of 
the Cosmic Lens All Sky Survey Radio-Selected Gravitational Lens Statistics}

\author{K.-H.~Chae}
\affiliation{University of Manchester, Jodrell Bank Observatory,
Macclesfield, Cheshire, SK11 9DL, UK}
\author{A.D.~Biggs}
\affiliation{University of Manchester, Jodrell Bank Observatory,
Macclesfield, Cheshire, SK11 9DL, UK}
\author{R.D.~Blandford}
\affiliation{California Institute of Technology, Pasadena, CA 91125}
\author{I.W.A.~Browne}
\affiliation{University of Manchester, Jodrell Bank Observatory,
Macclesfield, Cheshire, SK11 9DL, UK}
\author{A.G.~de~Bruyn}
\affiliation{NFRA, Postbus 2, 7990 A A Dwingeloo, The Netherlands}
\author{C.D.~Fassnacht}
\affiliation{Space Telescope Science Institute, 3700 San Martin Drive, 
Baltimore, MD 21218}
\author{P.~Helbig}
\affiliation{University of Manchester, Jodrell Bank Observatory,
Macclesfield, Cheshire, SK11 9DL, UK}
\author{N.J.~Jackson}
\affiliation{University of Manchester, Jodrell Bank Observatory,
Macclesfield, Cheshire, SK11 9DL, UK}
\author{L.J.~King}
\affiliation{University of Bonn, Auf dem H\"{u}gel, 69, D-53121 Bonn, Germany}
\author{L.V.E.~Koopmans}
\affiliation{California Institute of Technology, Pasadena, CA 91125}
\author{S.~Mao}
\affiliation{University of Manchester, Jodrell Bank Observatory,
Macclesfield, Cheshire, SK11 9DL, UK}
\author{D.R.~Marlow}
\affiliation{Dept.\ of Physics and Astronomy, University of Pennsylvania, 
209 S. 33rd Street Philadelphia, PA 19104}
\author{J.P.~McKean} 
\affiliation{University of Manchester, Jodrell Bank Observatory,
Macclesfield, Cheshire, SK11 9DL, UK}
\author{S.T.~Myers}
\affiliation{National Radio Astronomy Observatory, P.O. Box 0, Soccoro, 
NM 87801}  
\author{M.~Norbury}
\affiliation{University of Manchester, Jodrell Bank Observatory,
Macclesfield, Cheshire, SK11 9DL, UK} 
\author{T.J.~Pearson} 
\affiliation{California Institute of Technology, Pasadena, CA 91125}
\author{P.M.~Phillips} 
\affiliation{University of Manchester, Jodrell Bank Observatory,
Macclesfield, Cheshire, SK11 9DL, UK}
\author{A.C.S.~Readhead} 
\affiliation{California Institute of Technology, Pasadena, CA 91125}
\author{D.~Rusin}
\affiliation{Harvard-Smithsonian Center for
Astrophysics, 60 Garden St., MS-51, Cambridge, MA 02138}
\author{C.M.~Sykes} 
\affiliation{University of Manchester, Jodrell Bank Observatory,
Macclesfield, Cheshire, SK11 9DL, UK}
\author{P.N.~Wilkinson}
\affiliation{University of Manchester, Jodrell Bank Observatory,
Macclesfield, Cheshire, SK11 9DL, UK}
\author{E.~Xanthopoulos}
\affiliation{University of Manchester, Jodrell Bank Observatory,
Macclesfield, Cheshire, SK11 9DL, UK}
\author{T.~York}
\affiliation{University of Manchester, Jodrell Bank Observatory,
Macclesfield, Cheshire, SK11 9DL, UK}

\date{\today}

\begin{abstract}
We derive constraints on cosmological parameters and
the properties of the lensing galaxies from
gravitational lens statistics based on the final Cosmic Lens All 
Sky Survey (CLASS) data. For a flat universe with a classical cosmological
constant, we find that the present matter fraction of the critical density is
$\Omega_{\rm m}=0.31^{+0.27}_{-0.14}$ (68\%) $^{+0.12}_{-0.10}$ (systematic).
For a flat universe with a constant equation of state for dark energy
$w = p_x({\mbox{pressure}})/\rho_x({\mbox{energy density}})$, 
we find $w < -0.55^{+0.18}_{-0.11}$~(68\%).
\end{abstract}



\maketitle

Gravitational lensing, in which a background source behind
a foreground object is seen as distorted 
and (de)magnified image(s), is well-understood by the simple physics of 
light deflections in a weak gravitational field. Hence, 
the analysis of lens statistics can provide a technique~\cite{Turneretal} 
for constraining cosmological parameters that is both powerful and
complementary to other methods~\cite{Cosmology}.
However, for lens statistics to be useful, 
it is vital to have an unbiased statistical
sample of sources that is complete within well-defined observational
selection criteria~\cite{CLASS1and2} and to properly take into account all
factors in a statistical lensing model~\cite{CLASS3}.
The statistical properties of gravitational lensing in a sample are 
the total lensing rate, the image separations, the lens redshifts, 
the source redshifts, and the image multiplicities.
These properties depend not only on cosmological parameters
but also on the properties of the lensing galaxies and the 
distributions of the sources in redshift ($z$) and in luminosity in
the observational selection waveband (e.g.\ radio here)~\cite{CLASS3}.

Recent advances in observations permit us to do a reliable analysis
of lens statistics to obtain limits on cosmological parameters.
First, the largest radio-selected galactic mass-scale gravitational lens 
search project to date, the Cosmic Lens All Sky Survey (CLASS), has now 
been completed, resulting in the largest sample suitable for statistical 
analyses~\cite{CLASS1and2}.  
Second, currently available independent data sets on the distributions of
(flat spectrum) radio sources in flux density and in redshift are 
concordant~\cite{CLASS1and2,CLASS3}.
Third, recent large-scale observations of galaxies, in particular the
Two Degree Field Galaxy Redshift Survey (2dFGRS) and the Sloan Digital Sky
Survey (SDSS), have produced converging results on the total galaxy luminosity
function (LF; i.e., distribution of galaxy number density in luminosity), 
and there exist fairly reliable observational results
that permit us to extract from the total LF morphological type-specific
LFs, i.e., an early-type (i.e.\ ellipticals and S0 galaxies) LF and a 
late-type (i.e.\ spirals and others) LF~\cite{CLASS3,LF}. 
In this Letter, we report the main results on cosmological parameters from
a likelihood analysis of lens statistics based on these crucial sets of data. 
The detailed procedure of the analysis will be described in a later 
publication~\cite{CLASS3} in which extended analyses will be presented for
broader range of parameters and with an emphasis on the global properties of 
galaxy populations.

The CLASS statistical sample suitable for our analysis
contains 8958 radio sources out of which 13 sources 
are multiply-imaged~\cite{CLASS1and2}. The lens search 
looked for multiple-lensing image components of the central compact 
radio core in the background sources. The CLASS statistical sample is defined
so as to meet the following observational selection criteria~\cite{CLASS1and2}:
(1) the spectral index between 1.4~GHz and 5~GHz is flatter than $-0.5$, 
i.e.\ $\alpha \ge -0.5$ with $S_\nu \propto \nu^{\alpha}$ where 
$S_\nu$ is flux density measured in milli-jansky (mJy; 
1~mJy~$\equiv$~$10^{-29}$~W~m$^{-2}$~Hz$^{-1}$); (2) the total flux
density of each source is $\ge$~30~mJy at 5~GHz; (3) the total flux density 
of each source is $\ge$~20~mJy at 8.4~GHz; 
(4) the image components in lens systems
must have separations $\ge 0.3$~arcsec and the ratio of the flux densities of 
the fainter to the brighter component in double-image systems must be 
$\ge 0.1$. Number counts for
sources at $\nu = 5$~GHz are consistent with a (broken) power-law differential
number-flux density relation $|dN/dS|$ (i.e.\ number of sources per flux 
density bin) $\propto (S/S^0)^{-\eta}$ with $\eta=2.07\pm 0.02$
($1.97\pm 0.14$) for $S \ge S^0$ ($S \le S^0$) and $S^0 = 30$~mJy. 
These results are consistent with the prediction of the Dunlop \& Peacock 
(1990) free-form model number 5 radio luminosity function 
that is consistent with redshift data at 
$S > 1$~mJy~\cite{CLASS1and2,CLASS3}. Redshift measurements for
a representative CLASS subsample and the prediction of the aforementioned
Dunlop \& Peacock (1990) model consistently suggest that 
the redshift distribution for the CLASS unlensed sources can be 
adequately described by a Gaussian model with mean redshift, 
$\langle z \rangle = 1.27$ and dispersion 0.95~\cite{CLASS1and2,CLASS3}. 
The properties of the 13 multiply-imaged sources are summarized in Table~1.
\begin{table}
\caption{Gravitational lens systems in the well-defined CLASS statistical
sample. $z_s$ and $z_l$ are source and lens redshifts, respectively;
for sources with unmeasured redshifts, we take $z_s = 2$, which is the mean
source redshift for the entire sample of CLASS lenses. The maximum image 
separation ($\Delta\theta$) is given in arcsec. Image multiplicity is listed
under N$_{\rm im}$. When the image splitting is due to multiple galaxies, the
image separation and the image multiplicity are not used in our model fitting. 
The image separation of 2045+265 is not used because of uncertainties
in the nature of the lensing scenario. Probable lens galaxy type identification
(G-type) is coded as: e for an early-type and s for a spiral-type.}
\begin{ruledtabular}
\begin{tabular}{llllll}
Source    &  $z_s$   & $z_l$ & $\Delta\theta$  
& N$_{\rm im}$  & G-type \\
\hline
0218+357    &  0.96  & 0.68  & 0.334  & 2  & s \\
0445+123    &  -     &  -     & 1.33    & 2  & - \\
0631+519    &  -     &  -     & 1.16   & 2  & - \\
0712+472    &  1.34  & 0.41  & 1.27   & 4    & e \\
0850+054    &   -    &  -    & 0.68   & 2  & - \\
1152+199     &  1.019 & 0.439 & 1.56   & 2  & - \\
1359+154     &  3.235 &  -    & 1.65    & 6    & - (3 Gs) \\
1422+231     &  3.62  & 0.34  & 1.28    & 4    & e \\
1608+656     &  1.39  & 0.64  & 2.08    & 4    & e (2 Gs) \\
1933+503     &  2.62  & 0.755 & 1.17    & 4    & e? \\
2045+265    &  - & 0.867  & 1.86    & 4  & - \\
2114+022     &  -  & 0.32/0.59 & 2.57   & 2?  & e (2 Gs) \\
2319+051    &  -     & 0.624 & 1.36    & 2 & e  \\
\end{tabular}
\end{ruledtabular}
\end{table}

The luminosity function of galaxies is assumed to be 
well-described by a Schechter function~\cite{LF} of the form
\begin{equation}
\frac{dn}{d(L/L_*)} = n_* 
\left(\frac{L}{L_*}\right)^{\alpha}  \exp(-L/L_*),
\end{equation} 
where $\alpha$ is a faint-end slope and $n_*$ and $L_*$ are characteristic
number density and characteristic luminosity, respectively. From {Eq.}~(1) 
an integrated luminosity density $j$ is given by
$j = \int_0^{\infty} dL L (dn/dL) = n_* L_* \Gamma(\alpha+2)$. The
luminosity of a galaxy is correlated with its line-of-sight stellar
velocity dispersion ($\sigma$) via the empirical relations,
\begin{equation}
\frac{L}{L_*} \equiv 10^{0.4(M_*-M)} = 
\left(\frac{\sigma}{\sigma_*}\right)^{\gamma},
\end{equation}
where $M_*$ and $\sigma_*$ are respectively characteristic absolute magnitude
and characteristic velocity dispersion, and $\gamma$ corresponds to the
Faber-Jackson (Tully-Fisher) exponent for an early-type (late-type) 
population~\cite{FJTF};
throughout we take $\gamma_{\rm Faber-Jackson} = 4.0$
and $\gamma_{\rm Tully-Fisher} = 2.9$~\cite{FJTF}. 
The 2dFGRS team and the SDSS team independently determine the Schechter
parameters for galaxies at $z \lesssim 0.2$ summed over all galactic
types and their results are consistent with each other~\cite{LF}.  
The converging result can be expressed in the
$b_J$ photometric system as $M_*-5\log_{10}h = -19.69\pm 0.04$ (hereafter
$h$ is the Hubble constant $H_0$ in units of 100~km~s$^{-1}$~Mpc$^{-1}$), 
$\alpha = -1.22\pm 0.02$,
and $n_* = (1.71\pm 0.07)\times 10^{-2}$~$h^3$~Mpc$^{-3}$, which 
gives an integrated luminosity density of $j = (2.02\pm 0.15)\times 10^8$
$h$~$L_{\odot}$~Mpc$^{-3}$. From the total LF we extract the type-specific LFs
based on measured {\it relative} type-specific LFs keeping the total
luminosity density fixed. 
The type-specific LFs are required because the early-type 
and the late-type populations are dynamically different and 
contribute to the multiple imaging in distinctively different
ways, namely that the early-type population contributes more to the 
total lensing rate and on average generates larger image 
splittings~\cite{CLASS3}.
We use the relative LFs obtained from 
the Second Southern Sky Redshift Survey (SSRS2), which is based on 5404 
local ($z \leq 0.05$) galaxies. 
We obtain the following early-type ($e$) and late-type ($s$) 
LFs expressed in the $b_J$ photometric system: for the early-type population, 
$M_*^{(e)}-5\log_{10}h = -19.63^{+0.10}_{-0.11}$,
$\alpha^{(e)} = -1.00\pm 0.09$, and
$n_*^{(e)} = (0.64\pm 0.19) \times 10^{-2}$~$h^3$~Mpc$^{-3}$; 
 for the late-type population,
$\alpha^{(s)} = -1.22\pm 0.02$, $M_*^{(s)} = -19.69\pm 0.04$, and
$n_*^{(s)} = (1.20\pm 0.11) \times 10^{-2}$~$h^3$~Mpc$^{-3}$. 
To break a degeneracy we have chosen $\alpha^{(s)}=\alpha$ 
and  $M_*^{(s)}=M_*$ since the total LF is 
dominated by late-type galaxies. For alternative type-specific LFs
obtained from the 2dFGRS, the reader is referred to~\cite{CLASS3,LF}. When the
alternative type-specific LFs are used, our derived limits on cosmological 
parameters are essentially unchanged.

The multiple imaging cross section of a galaxy can be related to and 
calculated from the line-of-sight velocity dispersion of the galaxy. 
However, the relation is not unique but depends on the mass profile 
and intrinsic shape of the galaxy. 
This dependence will be absorbed into a ``dynamical normalization'' 
factor $\lambda (f)$ below. It is known that projected surface densities of
the inner cylindrical regions of galaxies along the line-of-sight
probed by gravitational lensing are well approximated
by a singular isothermal mass profile or a profile
similar to it~\cite{CLASS3}. In this Letter, we adopt a singular 
isothermal ellipsoid (SIE) as a galaxy lens model, 
whose projected density can be expressed in units of
lensing critical surface density 
$\Sigma_{\rm cr} = c^2 D_{\rm OS}/(4 \pi G D_{\rm OL} D_{\rm LS}) $ as
\begin{equation}
\kappa (x,y)  = \frac{r_{\rm cr}}{2}
  \frac{\sqrt{f} \lambda(f)}{\sqrt{x^2 + f^2 y^2}},
\end{equation}
where $f$ is the apparent minor-to-major axis ratio, $\lambda(f)$ is
the dynamical normalization factor for a given dynamical model, and
\begin{equation}
r_{\rm cr} = 4\pi 
\left(\frac{\sigma}{c}\right)^2
 \frac{D_{\rm OL} D_{\rm LS}}{D_{\rm OS}}
\end{equation}
is the critical radius for $f = 1$. $D_{\rm OL}$, $D_{\rm LS}$ and 
$D_{\rm OS}$ are angular diameter distances between observer (O), lensing 
galaxy (L), and source (S). Calculations of dynamical normalizations for
axisymmetric cases (i.e.\ oblate and prolate) can be found in~\cite{CLASS3}.
For the case of equal numbers of oblates and prolates, 
$|\lambda(f) - 1| \lesssim 0.06$ for $f > 0.4$.

Let $s_m (f)$ be the lensing cross section for image multiplicity $m$ 
($=2, 4$) due to the surface density given by equation~(3) in units of 
$r_{\rm cr}^2$ ({Eq.}~4). 
Then, we have $s_m(f) = [\lambda(f)]^2 \hat{s}_m(f)$ where
$\hat{s}_m(f)$ can be evaluated numerically~\cite{CLASS3}. 
The differential probability for a source with redshift $z_s$ and 
flux density $S$ to be multiply-imaged with image 
multiplicity $m$ and image separation $\Delta\theta$ to 
$\Delta\theta+d(\Delta\theta)$, due to a distribution 
of intervening galaxies of type $g$ ($g = e$, $s$) at  
$z$ to $z+dz$ modeled by SIEs and described by a type-specific Schechter LF, 
is given by
\begin{eqnarray}
\frac{d^2p_m^{(g)}}{dz d(\Delta\theta)} & = & 
s_m(f) 8 \pi^2 \gamma 
 \left|c\frac{dt}{dz}\right|  (1+z)^3
\left( \frac{D_{\rm OL}D_{\rm LS}}{D_{\rm OS}} \right)^2  \nonumber \\
  & & \times n_* \left( \frac{\sigma_*}{c} \right)^4
\frac{1}{\Delta\theta_*} 
\left( \frac{\Delta\theta}{\Delta\theta_*} 
\right)^{(\alpha \gamma + \gamma + 2)/2} \nonumber \\
 & & \times \exp \left[ - \left(\frac{\Delta\theta}{\Delta\theta_*} 
\right)^{\gamma/2} \right] B_m(z_s, S), 
\end{eqnarray}
where $B_m(z_s, S)$ is a magnification bias factor (i.e., an 
over-representation of multiply-imaged sources because of flux amplifications)
which depends on a magnification probability distribution and 
$|dN/dS|$ and can be calculated numerically~\cite{CLASS3}. 
In {Eq.}~(5), $\Delta\theta_*$ is a characteristic image (angular) separation
given by
\begin{equation}
\Delta\theta_* = \lambda(f) 8 \pi \frac{D_{\rm LS}}{D_{\rm OS}} 
\left( \frac{\sigma_*}{c} \right)^2. 
\end{equation}
The differential probability with $\Delta\theta$ to 
$\Delta\theta+d(\Delta\theta)$ can be obtained from {Eq.}~(5):
\begin{equation}
\frac{dp_m^{(g)}}{d(\Delta\theta)} = \int_{0}^{z_s} 
\frac{d^2 p_m^{(g)}}{dz d(\Delta\theta)} dz.
\end{equation}
The differential probability  
with any  $\Delta\theta > \Delta\theta_{\mbox{\scriptsize min}}$ due to 
galaxies at $z$ to $z+dz$ is given by
\begin{eqnarray}
\frac{dp_m^{(g)}}{dz} & = & s_m(f) 16 \pi^2 
\Gamma\left(\alpha+1+\frac{4}{\gamma}\right)
 \left|c\frac{dt}{dz}\right| (1+z)^3    \nonumber \\ 
 &  &  \times n_{*} \left( \frac{\sigma_*}{c} \right)^4 
\left( \frac{D_{\rm OL}D_{\rm LS}}{D_{\rm OS}}\right)^2 B_m(z_s, S)\nonumber\\
  &  &  - \int_0^{\Delta\theta_{\mbox{\scriptsize min}}} 
\frac{d^2p_m^{(g)}}{dz d(\Delta\theta)} d(\Delta\theta).
\end{eqnarray}
Finally, the integrated probability is given by 
\begin{equation}
p_m^{(g)} = \int_{0}^{z_s} \frac{dp_m^{(g)}}{dz} dz.
\end{equation}

For the sample of CLASS sources containing $N_{\rm L}$ multiply-imaged sources
and $N_{\rm U}$ unlensed sources, the likelihood $\mathcal{L}$ of observing the
number and the properties of the multiply-imaged sources (Table~1) and the
number of the unlensed sources given the statistical lensing model,
is defined by
\begin{equation}
\ln {\mathcal{L}} = \sum_{i=1}^{N_{\rm U}} 
\ln \left[1 - \sum_{m=2,4} p_m^{({\rm all})}(i)\right] + 
 \sum_{k=1}^{N_{\rm L}} \ln \delta p_m^{({\rm one})}(k), 
\end{equation}
where the integrated probability ({Eq.}~9) 
$p_m^{({\rm all})}(i)=p_m^{(e)}(i)+p_m^{(s)}(i)$ 
and the differential probability
({Eqs.}~5, 7, or 8) $\delta p_m^{({\rm one})}(k)$ is given by
\begin{equation}
\delta p_m^{({\rm one})}(k) = 
w^{(e)}(k) \delta p_m^{(e)}(k) + w^{(s)}(k) \delta p_m^{(s)}(k)
\end{equation}
with $w^{(e)}(k) + w^{(s)}(k) = 1$.
If the $k$-th lensing galaxy type is known to be an early-type (late-type), 
$w^{(e)}(k) =1$ ($w^{(s)}(k) =1$). If the $k$-th lensing galaxy type is 
unknown, we use 
$w^{(g)}(k) = \delta p_m^{(g)}(k)/[\delta p_m^{(e)}(k)+\delta p_m^{(s)}(k)]$ 
 if the lens redshift is known, and
we use $w^{(e)}(k) = 0.8$ and $w^{(s)}(k) = 0.2$ otherwise. From {Eq.}~(10), 
a ``$\chi^2$'' is defined as
\begin{equation}
\chi^2 = -2 \ln {\mathcal L}.
\end{equation}

We calculate the $\chi^2$ ({Eq.}~12) over grids of the model free parameters
(see below; in each grid point the $\chi^2$ is minimized over the nuisance 
parameters)
and determine the best-fit values and errors of the free parameters.
We allow most of the critical parameters to be free while we fix
the parameters that are well-determined by observations or to which the 
$\chi^2$ is insensitive. The free parameters to be determined by minimizing
the $\chi^2$ are as follows: present matter density $\Omega_{\rm m}$, present
vacuum density $\Omega_{\Lambda}$ or dark energy density $\Omega_x$ (all 
expressed as fractions of the present critical density) with its constant 
equation of state $w=p_x/\rho_x$ (where $p_x$ and $\rho_x$ are its isotropic
pressure and uniform energy density), 
early-type and late-type characteristic velocity 
dispersions $\sigma_*^{(e)}$ and $\sigma_*^{(s)}$, and the mean apparent axial
ratio of galaxies $\bar{f}$. A critical but fixed parameter is the early-type 
characteristic number density $n_*^{(e)}$. However, whichever of the presently
available choices of type-specific LFs we use, the derived constraints on 
cosmological parameters turn out to be virtually the same. One key assumption 
we make is that the (relatively) locally determined early-type number density
$n_*^{(e)}$ along with the faint-end slope $\alpha^{(e)}$ have not evolved
since $z \sim 1$. This appears to be a valid assumption 
based on current evidence and understanding~\cite{Peebles}. 
However, if there is any significant reduction in the early-type number
density at $z \sim 0.6$ compared with the present epoch,
our presently determined matter density (dark energy density) will
be overestimated (underestimated).

We present our results based on the type-specific LFs 
derived using the SSRS2 LFs and a dynamical
normalization assuming equal frequencies of oblates and prolates.
\begin{figure}[t]
\begin{center}
\setlength{\unitlength}{1cm}
\begin{picture}(6.6,6.6)(0,0)
\put(-1.,-2.){\includegraphics{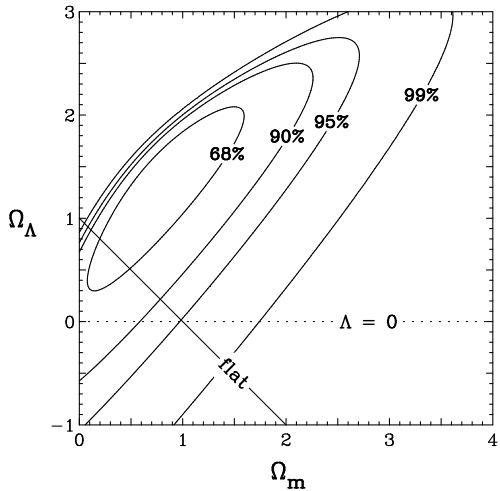}}
\end{picture}
\caption{Confidence limits in the $\Omega_{\rm m}$-$\Omega_{\Lambda}$ plane.}
\label{}
\end{center}
\end{figure}
Figure~1 shows likelihood regions in a matter plus vacuum cosmology: 
we find the combination 
$\Omega_{\Lambda}-1.2\Omega_{\rm m}=0.40^{+0.33}_{-0.56}$ 
(68\%) ($^{+0.52}_{-1.57}$ at 95\%). 
Figure~2 shows the likelihood in the $w$-$\Omega_{\rm m}$ plane with 
the prior that the universe is flat; 
we find $w < -0.55_{-0.11}^{+0.18}$ (68\%).
This limit is in agreement with other recent results~\cite{Eqofstate}. 
For a flat universe with a cosmological constant (i.e.\ for $w = -1$),
we find $\Omega_{\rm m}=0.31_{-0.14}^{+0.27}$ (68\%)
($_{-0.23}^{+0.75}$ at 95\%) $_{-0.10}^{+0.12}$ (systematic).
The identified systematics are due to 
the uncertainties in the redshift distribution 
and the differential number-flux density relation of the sources. 
For the flat $w=-1$ case, other fitted values of the parameters are: 
$\sigma_*^{(e)}= 198^{+53}_{-37}$~km~s$^{-1}$, 
$\sigma_*^{(s)}= 117^{+45}_{-31}$~km~s$^{-1}$, 
and $\bar{f} < 0.83$, all at 95\%. 

In conclusion, the likelihood function (or, $\chi^2$; {Eq.}~12) of the
CLASS data has well-defined confidence regions in the parameter space 
defined by our cosmological and galactic parameters. 
Our constraint in the $\Omega_{\rm m}$-$\Omega_{\Lambda}$ plane based 
solely on lens statistics agrees with that from 
Type~Ia supernovae observations~\cite{Cosmology}.
Our determined value of $\Omega_{\rm m}$ in flat cosmologies 
is also in good agreement with several recent independent 
determinations~\cite{Cosmology}. Compared with previous works in lens 
statistics~\cite{CLASS3}, our results imply not only positive  
but higher dark energy density. 
Therefore, our results, which are based on 
different physics, assumptions and astronomical relations, 
and independent sets of data, add to the evidence 
for a universe with low matter density and high dark-energy density.

\newpage
\begin{figure}[t]
\begin{center}
\setlength{\unitlength}{1cm}
\begin{picture}(6.6,6.6)(0,0)
\put(-1.,-2.){\includegraphics{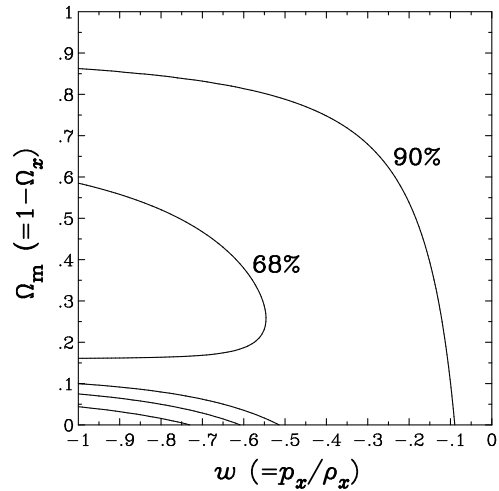}}
\end{picture}
\caption{Confidence limits in the $w$-$\Omega_{\rm m}$ plane in flat 
cosmologies.}
\label{}
\end{center}
\end{figure}

\end{document}